# Countering Wrapping Attack on XML Signature in SOAP Message for Cloud Computing

Hadi Razzaghi Kouchaksaraei, Alexander G. Chefranov

*Department of Computer Engineering, Eastern Mediterranean University*
*Gazimagusa- Mersin 10- Turkey*
Hadi.razzaghi87@gmail.com
Alexander.chefranov@emu.edu.tr

*Abstract*— **It is known that the exchange of information between web applications is done by means of the SOAP protocol. Securing this protocol is obviously a vital issue for any computer network. However, when it comes to cloud computing systems, the sensitivity of this issue rises, as the clients of system, release their data to the cloud.**

**XML signature is employed to secure SOAP messages. However, there are also some weak points that have been identified, named as XML signature wrapping attacks, which have been categorized into four major groups; Simple Ancestry Context Attack, Optional element context attacks, Sibling Value Context Attack, Sibling Order Context.**

**In this paper, two existing methods, for referencing the signed part of SOAP Message, named as ID referencing and XPath method, are analyzed and examined. In addition, a new method is proposed and tested, to secure the SOAP message.**

**In the new method, the XML any signature wrapping attack is prevented by employing the concept of XML digital signature on the SOAP message. The results of conducted experiments show that the proposed method is approximately three times faster than the XPath method and even a little faster than ID.**

*Keywords: Cloud Computing, SOAP message, XML digital signature, Wrapping attack.*

## I. INTRODUCTION

Cloud computing is a new technology [1], which provides greatly ascendable resources such as bandwidth, hardware and software, to be utilized as a service for consumers, over the Internet. This concept has attracted wide attention in all kind of industries recently [2]. One of the most significant advantages of using of this technology is that consumers can save the cost of hardware deployment, software license and system maintenance. Consequently, the price of providing and using the systems will be reduced significantly.

However, besides being absolutely beneficial, there are still particular unsolved problems [2], in order to implement this concept. It can be said that the most important challenges in cloud computing are security and trust. Since the consumer's data has to be released to the cloud, the system requires high security and safety over them. The data in clouds could be very personal and sensitive and must not be unveiled to an unauthorized person. In cloud computing, data are threatened during the transition as well. This problem reduces the reliability of the cloud systems [3].

A popular protocol, which is used to exchange the data in cloud systems, is Simple Object Access Protocol (SOAP) [4] based on Extensible Markup Language (XML) [5]. Securing data in SOAP messages is one of the main concerns related to security in cloud systems. It can be threatened by XML Signature wrapping attack, which causes the unveiling of sensitive data [6]. This attack is based on altering the structure of the original message from the genuine sender. Although some remedies have been proposed to counter this attack (ID referencing and XPath methods), none of them has been able to counter the attack completely [6], as they sign a particular part of an XML document.

The solution provided in this research, uses a new method, namely SESoap, to provide integrity for the messages exchanged in a cloud system by SOAP. In this technique, which is less complicated, more reliable and faster than the ID referencing and XPath methods, the entire SOAP message is signed by XML digital signature, instead of signing a part of that. It also counters all known wrapping attacks and makes similar attacks impossible.

Layout of this article is as follows. In the next section, basic definitions and explanations related to SOAP message and XML signature are given. In the 3rd section, XML signature wrapping attack and its four different categories are explained briefly. The 4th section covers some of the previous researches, which are relevant to this topic. Proposing and describing the SESoap method, its analysis, and their results have been given in the 5th section. Finally brief conclusions and achievements of this research have been given in 6th section.

## II. TECHNICAL BACKGROUND

### A. SOAP Message

SOAP [7], is a protocol to provide communication between applications. It works as a format for sending messages via Internet and also collaborates with the firewalls [8], [9], [7].

*1) SOAP Building Blocks:* As it is also mentioned above, SOAP message's language is based on XML [8]. Moreover, it can be explained that the building block of SOAP is in fact a typical XML document, which consists of these items:

1) Envelope: this element recognizes the XML document as a SOAP message.
2) Header: this element includes the header information of a SOAP message.
3) Body: this element includes the actual SOAP message
4) Fault: Errors that occurred while processing message are included in this element [8], [10].

*3) Skeleton of a SOAP Message:* A typical skeleton of a SOAP message is shown in Fig.1.

```
<?xml version="1.0"?>
<soap:Envelope
xmlns:soap="http://www.w3.org/2001/12/soap-envelope"
soap:encodingStyle="http://www.w3.org/2001/12/soap-encoding">

<soap:Header>
...
</soap:Header>

<soap:Body>
...
    <soap:Fault>
    ...
    </soap:Fault>
</soap:Body>

</soap:Envelope>
```

Fig. 1. Skeleton of a SOAP message [10]

### B. XML signature

XML signature is a technique, which is used to deliver reliability, integrity and message authentication, for various types of data [11]. By providing integrity to data, it is meant that once the data is signed; it cannot be altered later, without invalidating the signature. This technique is executed by employing asymmetric cryptography. The roles for signing a document are as follows [12].

$$M = D_{P_c}[E_{R_c}[M]] = D_{R_c}[E_{P_c}[M]]$$

In the formula, a message M is signed by private key and a public key is used to verify the signature. The reverse operation is allowed as well. Asymmetric encryption uses two keys in order to encrypt and decrypt a message, M, which are named private (Rc) and public (Pc) keys. XML digital signature employs private key and public key to sign a message and validate the document, respectively. When signing the message, signature will be attached to the original document, and will be sent to the receiver. It should be noted that the document, is not hidden, since hiding the message is not the aim of XML digital signature. Since asymmetric encryption is time consuming, a hash function (f (M)) is calculated over the document and the result, which is called digest value, is considerably smaller than the document itself. The result of hash function is then encrypted by private key. Consequently, the time passed for encrypting data is reduced significantly. Fig. 2 shows the structure of an XML signature.

```
<Signature>
    <SignedInfo>
        <CanonicalizationMethod />
        <SignatureMethod />
        <Reference>
            <Transforms>
            <DigestMethod>
            <DigestValue>
        </Reference>
        <Reference /> etc.
    </SignedInfo>
    <SignatureValue />
    <KeyInfo />
    <Object />
</Signature>
```

Fig. 2. Structure of an XML signature [12]

### III. XML SIGNATURE WRAPPING ATTACK

XML signature wrapping attacks are possible because of the fact that the signature does not convey any information to where the referenced element is placed [13]. This attack was introduced for the first time, in 2005 by McIntosh and Austel [14], stating different kind of this attack, including Simple Context, Optional Element, Optional Element in security header (sibling value) and Namespace injection (Sibling order) [14]. This attack happens in SOAP message, which transfers the XML document, over the Internet.

### A. Simple Ancestry Context Attack

In Simple Ancestry Context Attack, a request's SOAP body is signed by a signature, which is placed in the security header of the request. The recipient of the message, checks if the signature is correct and legalizes trust in the signing credential. Lastly, the recipient controls to realize whether the required element was actually signed, by bringing the "id" of the SOAP body to the ID reference, in the signature [15].

A typical example of this attack is shown in Fig. 3. The mechanism of this attack be briefly explained in this way that, the SOAP body gets swapped with a malicious SOAP body. The original SOAP body is placed in a <wrapper> element, which is situated in the SOAP header and when the signature is validated, the XML signature confirmation algorithm, begins searching for the element, which has the id of "CMPE", as it is stated in the <Reference> element. Finally, <soap:Header> Element wrapped within the <wrapper> element, will be found by the algorithm. The verification will be implemented on the <soap:Header>, within the <wrapper> element. The verification will be positive, because it includes the original SOAP body, which is signed by the sender. The SOAP message will be passed to the logic of the application. In the application logic procedure, only the SOAP body, which is straightly positioned under the SOAP header, will be processed. In other words, all other SOAP body elements will be just ignored [15]. Fig. 3 shows how this attack works.

```
<soap:Envelope ...>
    <soap:Header>
        <wsse:Security>
            ...
            <ds:Signature>
                ...
                <ds:Reference URI="#CMPE">
                ...
                </ds:Reference>
            </ds:SignedInfo>

            </ds:Signature>
        </wsse:Security>

        <Wrapper
        soap:mustUnderstand="0"
        soap:role="...none" >
        <soap:Body wsu:Id="CMPE">
            <getQute Symbol="IBM">
        </soap:Body>
        </Wrapper>
    </soap:Header>
    <soap:Body wsu:Id="newCMPE">
        <getQuote Symbol="MBI">
    </soap:Body>
</soap:Envelope>
```

Fig. 3. Typical Simple Ancestry Context Attack [16]

## B. Optional Element Context Attacks

In Optional Element Context Attacks, the signed data is contained in the SOAP header and it is arbitrary. Comparing this attack to the Simple Context Attack, which is explained above, reveals that the main problem is not the place of the signed data in the SOAP header [9]. In fact, the optional nature of signed data is the main issue [14]. The <ReplyTo> element, which specifies where to send the reply, can be given as an example, which is shown in Fig. 4. The mechanism of this attack can be explained as follows; it can be seen that the element of <wsa:ReplyTo> is placed in the <wrapper> element, while, the element of <wrapper> is also positioned underneath the <wsse:security>. In addition, by means of soap:mustUnderstand="0", in <wrapper>, this element has become optional and by using soap:role="../none", it is destined that the SOAP node (application logic) should not process this header element. These modifications in the SOAP message, result in the <wsa:ReplyTo> to become completely disregarded by the application's logic. Having these attributions, when the signature gets legalized, the verification algorithm of XML signature begins to search for the element, which has the id of "theReplyTo" (specified in the <Reference>) and <wsa:ReplyTo>, which is in the <wrapper> element, will be found. At this stage, signature confirmation will be done on the <wsa:ReplyTo>, in the <wrapper>, and because it is including the original <wsa:ReplyTo>, signature confirmation will be positive. Consequently, SOAP message body and the descendants, which are understood, will be handed to the application logic while the <wrapper>, will not be passed to it. Thus, the application logic will ignore the <wsa:ReplyTo> element and as the result, the reply will not go to the address specified in <wsa:ReplyTo> and the original message sender will get the reply [9].

## C. Sibling Value Context Attack

Sibling Value Context Attack covers the following scenario. In this attack, the security header includes a signed element, which is in fact an alternative sibling of <Signature>. A common model for this attack can be the element of <Timestamp>, which together with <Signature>, are direct descendants of SOAP security header. The difference between this attack and the previously discussed attacks is in the signed data, which in this attack is the sibling of <Signature> [16]. The main aim of this attack is to ignore the sibling of the signature element.

## D. Sibling Order Context

According to McIntoch and Austel, 2005[14], this attack is dealing with the protection of the sibling elements that are individually signed.

Their semantics are related to their order relative to one another, from reordering by an adversary. More work is required to define appropriate countermeasures that do not prevent the addition and removal of siblings that do not impact the ordering semantics [14].

## IV. KNOWN COUNTERMEASURES TO WRAPPING ATTACKS

The requirements of a service-side security policy, in order to detect an attack were shown by McIntosh and Austell, 2005 [14]. These necessities are being improved by each attack, which is able to bypass the previous provided security policy. In continuance, some of the improvements in the policy will be explained.

```
<soap:Envelope ...>
 <soap:Header>
  <wsse: Security>
   ...
   <ds:Signature>
    <ds:SignedInfo>
     ...
     <ds:Reference URI="#CMPE">
     ...
     </ds:Reference>
     <ds:Reference URI="#theReplyTo">
     ...
     </ds:Reference>
    </ds:SignedInfo>
    ...
   </ds:Signature>
  </wsse:Security>

        <Wrapper
   soap:mustUnderstand="0"
   soap:role="…/none" >
   <wsa:ReplyTo wsu:Id="theReplyTo">
    <wsa:Address>http://cmpe.emu.edu.tr/</wsa:Address>
   </wsa:ReplyTo>
  </Wrapper>

 </soap:Header>
 <soap:Body wsu:Id="CMPE">
  <getQuote Symbol="IBM"/>
 </soap:Body>
</soap:Envelope>
```

Fig. 4. Typical Optional element context attack [14]

1) In the wsse:security header element, a signature "A", XML signature, should be placed, having a clear soap:role attribute and value of "…/ultimateReceiver".

2) From signature "A", the element, identified by /soap:Envelope/soap:Body, must be referenced.

3) In the case of having any elements, which are matching with

/soap:envelop/soap:Header/wsse:Security[@role="…/ultimateReceiver"]   wsu:Timestamp   and /soap:Envelop/soap:Header/wsa:ReplyTo, it should be noted that these elements must be referred through an absolute path, Xpath expression, from signature "A".

4) Verification key of signature "A" must be issued and provided by a trusted Certificate Authorities (CAs) and the certificate of X.509v3, respectively [14].

The first example of XML signature wrapping attack, which was indicating that the controls suggested by McIntosh and Austell [14] are not satisfactory to notice XML signature wrapping attack, was shown by Gruschka and Lo Iacono, in 2009 [17]. It is also claimed in their research that the timestamp has to be referenced by an extra XPath expression, which is not fulfilled in Fig. 4. Although, it can be added easily, it should be noted that the XPath references result in further problems. It is known that XPath expressions are more difficult to be evaluated, comparing to IDs, this issue is especially important in the context of streaming SOAP message. Another more important issue is that employment of XPath references may indicate security issues, so they are not suggested by basic security profile [6].

In a new method [18], which was proposed in 2006 and is named as inline method, a new element called SOAP account was introduced. Some characteristic information are gathered together and inserted in the SOAP account element [17]. Protection of some key features of SOAP message structure is aimed in this technique. The properties, which are aimed to be protected, are listed as below.

1) Number of header element descendants
2) Number of soap:envelop, descendent elements
3) Amount of references in every signature
4) The descendants and antecedents of every signed item

By means of this approach, with the above properties, if in an attack, each of these properties is changed, the attack will be easily identified [18].

The main problem with this method is that it does not provide a general protection, from XML signature wrapping attack. In other words, if an attacker manages to change the SOAP message structure in a way that the inline method structure properties does not get changed, this technique can be easily dodged [19].

In addition, fastXPath method was proposed by Gajek et al., in 2009 [9]. This method is employed to increase the speed of XPath function, and to point to the signed subtree. However, this method also could not solve the identified issues about XPath expression [20]. A comparison between runtime of different methods, ID, fastXPath and XPath methods, have been also done in their investigation. The comparison's relevant graph is shown in Fig. 5.

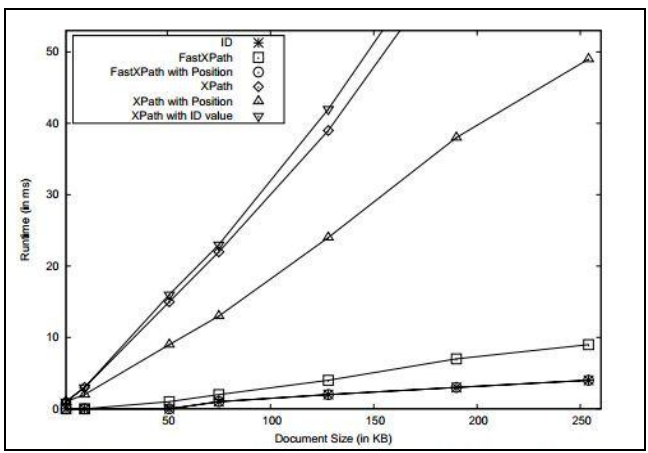

Fig. 5. Runtime comparison of different referencing methods [17]

## V. SIGNING ENTIRE SOAP (SESOAP) METHOD

Since most of the XML signature wrapping attacks are done through changing the structure of the original SOAP message, sent by the genuine sender [17], it is logical to propose a protecting method, which aims to protect the structure of the sent message, from attacker. To fulfil this aim, the digital signature can be used to guarantee the integrity of message.

The method of this paper, i.e. Signing Entire SOAP (SESoap) method, is to apply the digital signature structure over entire SOAP envelop element, which results in securing the whole document. Consequently, an attacker will not be able to change the location of elements or remove or add any element to the original document. In the case of modification in any part of the document, the signature cannot be verified. The skeleton of SESoap method is shown in Fig. 7.

It should be noted that the element of SOAP:signature, contains the result of signing the entire content of soap:envelop, except the element of soap:signature itself. To explain better, the structure of SOAP after applying the SESoap method is shown in Fig. 6.

### A. Simple Element Context Attack Countering

In simple Context attack, a wrapper alters the location of the Soap body and adds a new Soap body to threaten the SOAP document [14]. It is quite clear that by using digital signature over entire document, any alteration or adding any element to the signed document will be totally prevented.

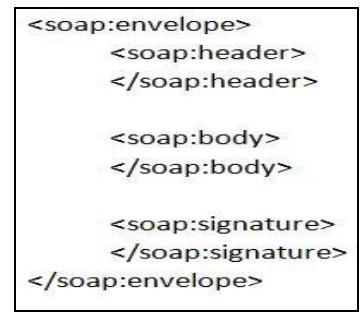

Fig. 6. Skeleton of SESoap method

### B. Optional Element Context Attack Countering

In Optional Element Context attack, a wrapper adds some information to optional element to application logic of a program could not parse that element [14]. Again, the same as the previous attack, when a wrapper tends to add something to the document, the attack is prevented by SESoap.

### C. Sibling Value Context Attack Countering

The two previous types of attacks are possible to be prevented by means of XPath method [14]; however XPath is susceptible against this attack [6]. As it has been explained in the previous section, Time stamp element, which is an optional sibling element of signature element, can be threatened by wrapper. But for wrapping on this element, the wrapper again must modify some parts of document [14]. Consequently, as modifications are prevented in SESoap method, Sibling Value Context attack will not be allowed to occur.

### D. Sibling Order Attack Countering

This attack relies on changing the order of individual sibling elements [14]. Therefore, since reordering is also not possible in SESoap, again no wrapper can be successful in implementation of this attack.

### E. Conducted Experiments

SESoap method has been implemented by using C#.net, in order to determine how fast it is, comparing to the previous methods of ID referencing and XPath. These examinations have been performed by means of Laptop, having 2.00 GHz Core2Duo CPU, and 1.00 GB memory, in Windows7 operating system. The time for finding the element in SESoap is zero, because this technique does not search for any specified element inside the SOAP document. Experiments were conducted on file sizes used in [17] and also on more than ten times greater size (up to 3.15 MB). The graph for comparing these time durations is shown in Fig. 7.

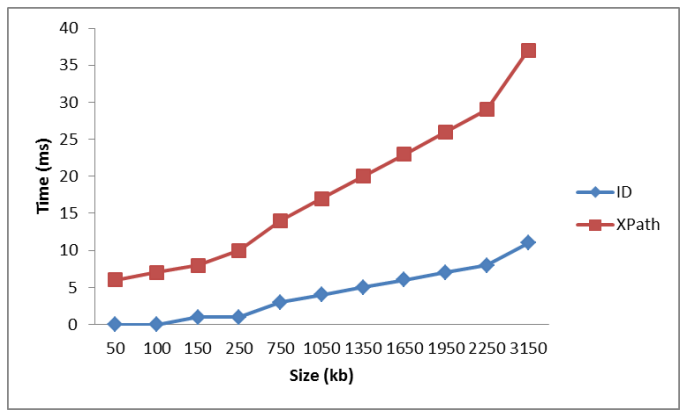

Fig. 7. Time durations of the ID and XPath methods

In the next step, the time durations for hashing the specified element inside the SOAP document, have been estimated. Fig. 8 shows the result of the hashing specified element.

In addition, the consumed time for encrypting data, in all the three methods are the same, because in the digital signature, encryption function applies on the signed info element of signature. The sizes of the signed info element in all the methods are equal. As the result, the consumed times for encrypting the signed info elements are the same. In this study the time consumed for all three methods was 3.0004 milliseconds. In this study, two codes have been used to measure the time, in Code1 each function (Finding element, hash function and encryption function) has been done separately and in Code2 the whole operations have been done as one component. The total times consumed to sign the soap message in each of the three methods, using Code1, are shown in Fig. 9 [21].

According to these results, ID is faster than XPath, in finding an element. On the other hand, ID and XPath methods are faster, comparing to SESoap method, in hashing the specified element. Moreover, as the numbers show, the total consumed time to sign a SOAP document by SESoap method is approximately three times faster than the XPath method and even a little faster than ID. Consequently, it can be claimed that, the SESoap method is operating more sufficiently, than the other two methods, considering both aspects of security and time.

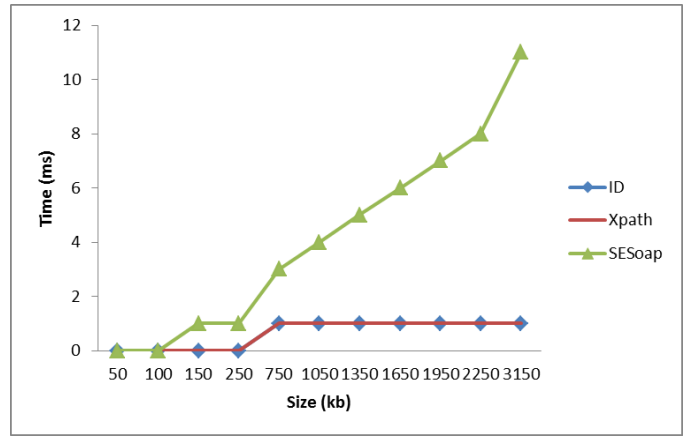

Fig. 8. Time durations for hashing the specified element

Moreover, the total time durations in order to sign soap message, using Code 2, is shown in Fig. 10.

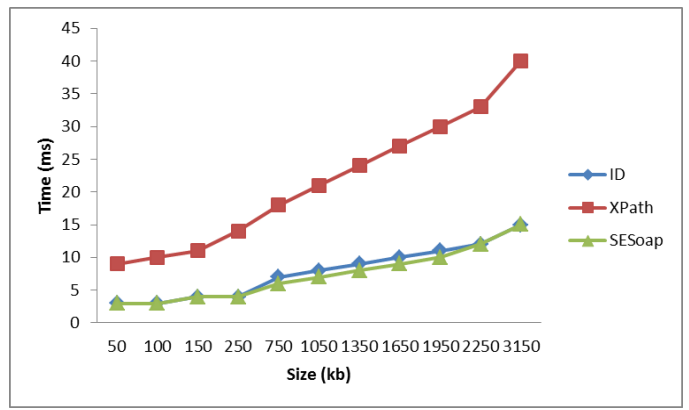

Fig. 9 Total time durations consumed to sign the soap message, using Code 1

These results are more complying with the previous research [17], but as it can be obviously noticed, the results of that research are less efficient than what is done in this study.

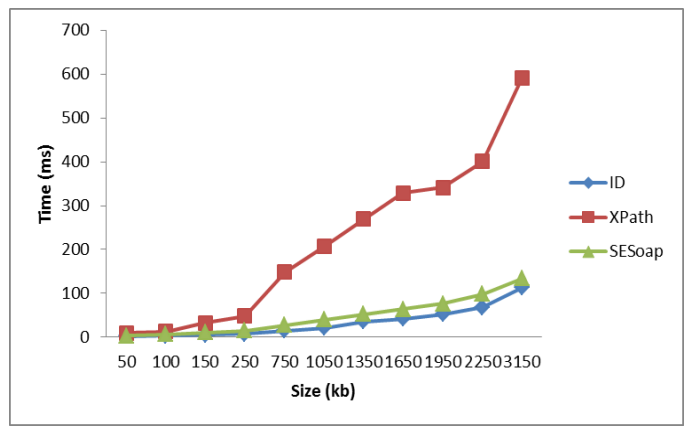

Fig. 10 Total time durations consumed to sign the SOAP message, using Code 2

## VI. Conclusion

The primary goal of this study was to secure SOAP message, which is employed to exchange information between web applications of cloud computing systems. Having this aim, a new method, SESoap, has been proposed. The concept of this method is using Digital Signature technique to immune the information inside a SOAP message from modification by an adversary.

The results obtained from implementation of SESoap method indicate that this method is slower than the other examined methods, for hashing the information. The reason of this observation is that, comparing to the other examined methods, in this method, the hash function is applied over a greater size of data. On the other hand, for finding element in SOAP message, SESoap does not consume any time and the total time duration for signing the message, is approximately three times faster than the XPath method and even a little faster than ID.